\documentclass[11pt,a4paper]{article}
\usepackage{graphicx} %
\usepackage{amsmath}
\usepackage{lineno}

\usepackage{relsize}
\usepackage{graphicx}
\usepackage{tcolorbox}
\usepackage{todonotes}
\usepackage[colorlinks=true,
allcolors=blue,
            pdfborder={0 0 0},
            ]{hyperref}

\title{US National Input to\\the European Strategy Update for Particle Physics}
\author{Executive Committees\\Division of Particles and Fields and Division of Physics of Beams\\American Physical Society\thanks{Corresponding authors: Andr\'e de Gouv\^ea (past-chair DPF), Hitoshi Murayama (vice-chair DPF, P5) , Mark Palmer (chair-elect DPB, P5), Heidi Schellman (chair, DPF)}}

\begin{document}

\maketitle
\thispagestyle{empty}
\setlength\itemsep{-3em}
\begin{abstract}
    In this document we summarize the output of the US community planning exercises for particle physics that were performed between 2020 and 2023 and comment upon progress made since then towards our common scientific goals. This document leans heavily on the formal report of the Particle Physics Project Prioritization Panel and other recent US planning documents, often quoting them verbatim to retain the community consensus. \end{abstract}

\pagebreak
\setcounter{page}{1}
\begin{center}
    {\bf \large US National Input to the European Strategy Update for Particle Physics}
\end{center} 
\begin{center}
\includegraphics[width=.9 \textwidth]{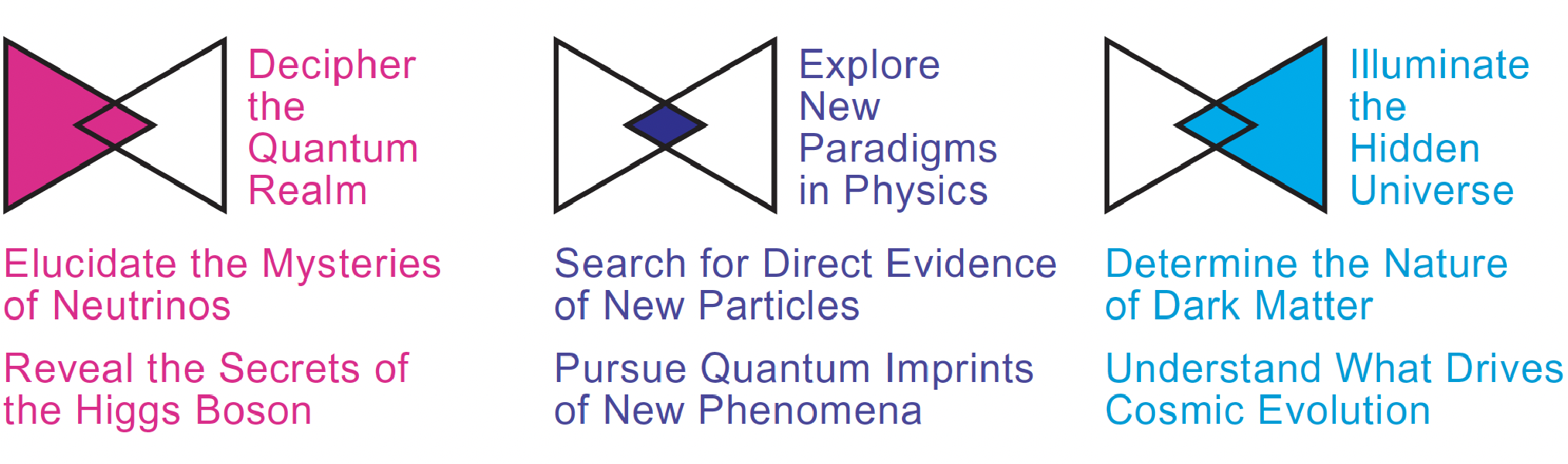}
\end{center}

The US particle physics community engages in periodic planning exercises. Since 2003, these have been formalized through the Snowmass and Particle Physics Project Prioritization Panel (P5) processes outlined below. This document describes the outcome from the most recent Snowmass/P5 planning process together with relevant related activity in Nuclear Science with emphasis on their international context. 
The P5 process was a true prioritization and only retained efforts that were consistent with reasonable funding projections. Although the panel was convened by the funding agencies, the final recommendations  were endorsed by over 3,200 members of the US particle physics community. 

The 2023 \href{https://www.usparticlephysics.org/2023-p5-report/}{Particle Physics Project Prioritization Panel} (P5) was charged by the National Science Foundation (NSF) and the Department of Energy (DOE) Office of Science with developing a 10-year strategic plan for US particle physics, in the context of a 20-year global strategy and two constrained budget scenarios. P5 formally reports to the High-Energy Physics Advisory Panel (HEPAP) appointed by DOE and NSF. 

In preparation for the development of a 10-year strategic plan for particle physics in the US, the \href{https://inspirehep.net/literature/2623768}{2021 Snowmass Community Planning Exercise} was organized by the Division of Particles and Fields (DPF) of the American Physical Society. 
The Snowmass process ran from 2019 to 2022. Over 500 white papers were submitted. The 2-week meeting in July 2022, in Seattle, WA had 700 in-person  and 650 remote participants. 

P5  gathered additional input from several other sources, including town hall meetings, laboratory visits, and individual communications.

The DOE provided the panel with two budget scenarios for High Energy Physics (HEP)
derived from realistic near-term budget projections. The baseline scenario assumes budget
levels for HEP for fiscal years 2023 through 2027 that are specified in the Creating Helpful
Incentives to Produce Semiconductors (CHIPS) and Science Act of 2022. The baseline
budget scenario then increases by 3\% per year from fiscal year 2028 through 2033. The
less favorable scenario assumes increases of 2\% per year from fiscal year 2024 to 2033.
The panel was asked to develop compelling science programs consistent with these scenarios. %

The community presented P5 with more inspiring and ambitious projects than either budget scenario could accommodate. To guide the necessary choices, the panel categorized projects as small, medium, and large, based on their construction costs to the particle physics program. In the large and medium categories, initiatives were first prioritized based on individual scientific merit, then design maturity. An expert subcommittee independently reviewed the costs, technical risks, and schedule of large projects. The final prioritization holistically considered the cost of construction, commissioning, operations, and related research support, distributed over a 10- to 20-year period.

The P5 report was approved by HEPAP on December 8, 2023. DPF \href{https://docs.google.com/spreadsheets/d/e/2PACX-1vRaGc9kVV-uBNPJstmYcYMwiluxya7hSsgDnWB1pg44l8zxPvGl4jUSvOR0VCoa5fCetgDJCNDUJb3O/pubhtml?gid=307089066&single=true}{collected} over 3,600  endorsements from the global community, of which more than 3,200 came from US scientists, demonstrating  broad community support for the P5 recommendations. 

In drafting this document, the corresponding authors consulted around 50 experts across particle, beam and nuclear physics and hosted an open \href{https://indico.global/event/13796/}{forum} on Feb. 27 2025, attended remotely by $\simeq 180$ participants where plans for white papers were presented. The paper was then opened up for general comment by the community. This consultation and a \href{https://indico.global/event/14242/timetable/}{collection of US white papers}, which are linked in the supporting documents, informed our responses.

\vspace{-1 em}
\subsection*{The P5 Recommendations}
\begin{figure}[th]
\begin{center}
\includegraphics[width=0.7\textwidth]{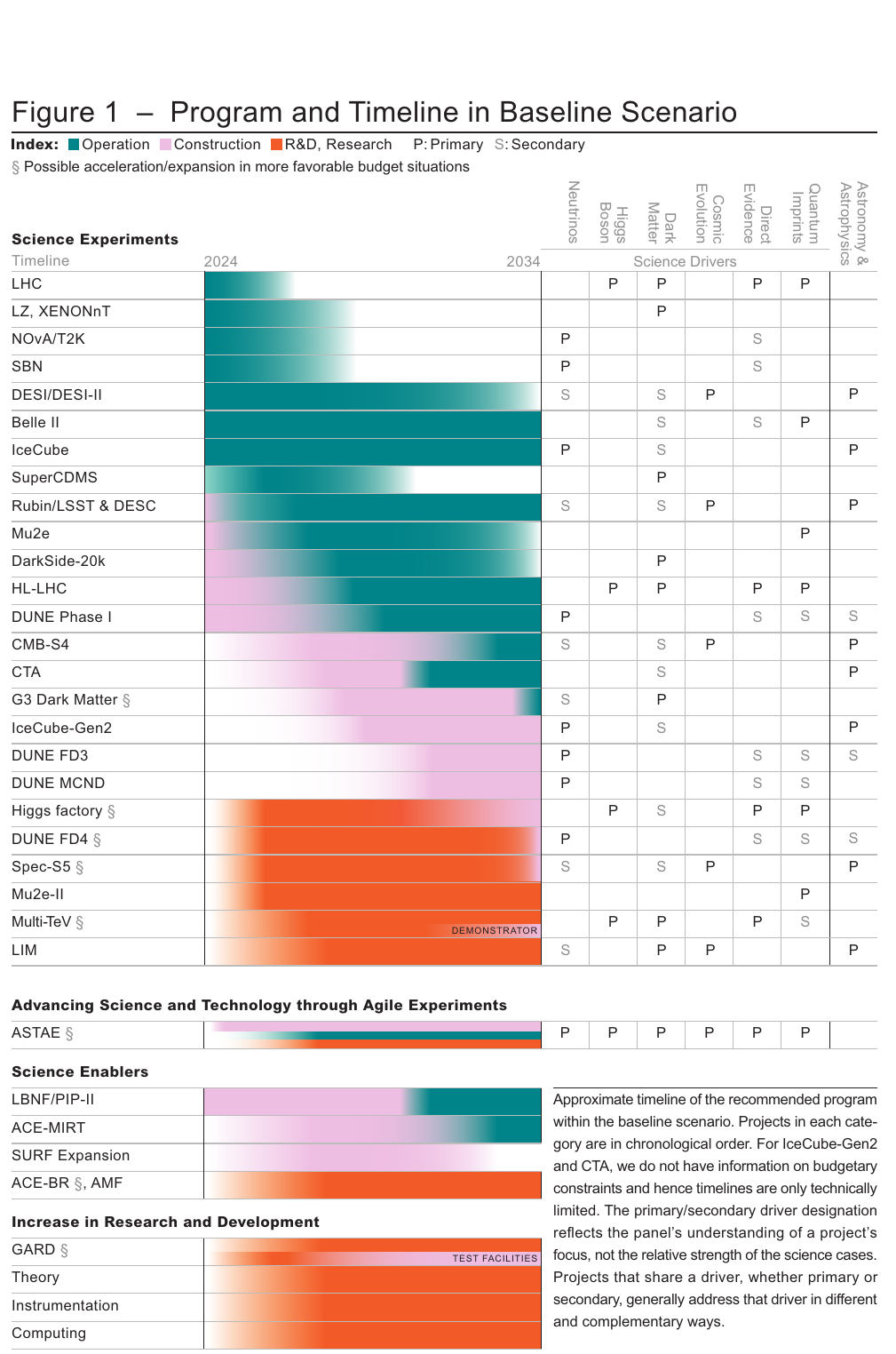}
\end{center}
\caption{Approximate timeline of the recommended program within the baseline scenario. Projects in each category are in chronological order. For IceCube-Gen2 and CTA, we do not have information on budgetary constraints and hence timelines are only technically limited. The primary/secondary driver designation reflects the panel’s understanding of a project’s focus, not the relative strength of the science cases. Projects that share a driver, whether primary or secondary, generally address that driver in different and complementary ways.}
\label{Fig1}
\end{figure}

\begin{figure}[th]
\includegraphics[width=0.85\textwidth]{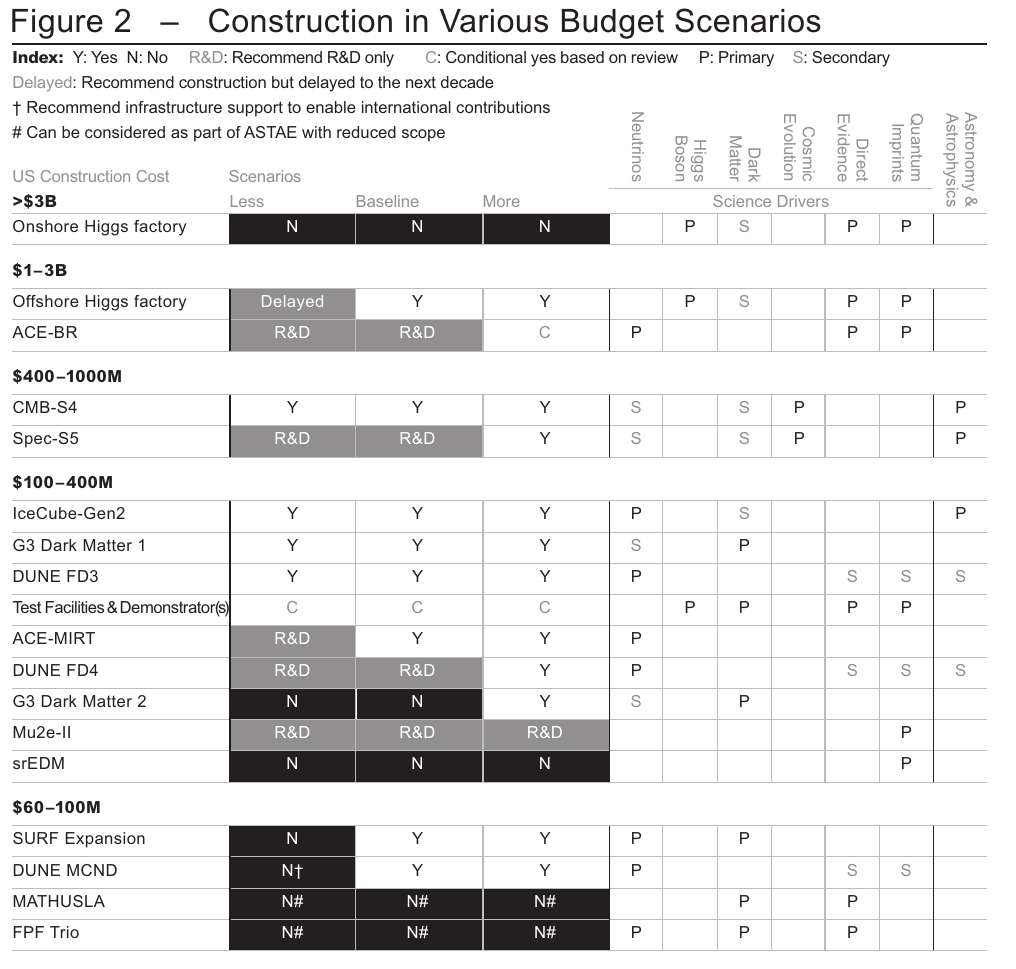}
\caption{Medium and large-scale US investments in
new construction projects for possible budget
scenarios. For the three budget scenarios,
the projects are ordered in 5 budget brackets
according to the number of “N” entries and
then by approximate budget sizes. For the
offshore Higgs factory, test facilities \& demonstrators,
see \protect\hyperlink{rec6}{Recommendation 6}. 
See the caption of Figure \ref{Fig1} concerning the science drivers, and \href{https://www.usparticlephysics.org/2023-p5-report/budgetary-considerations.html}{section 8} for the rationale behind these choices. Note that the cost brackets are provided in the US accounting method.}
\label{Fig2}
\end{figure}

For convenience, we gather here, verbatim, the primary recommendations from the \href{https://www.usparticlephysics.org/2023-p5-report/full-list-of-recommendations}{P5 report} that are relevant to the European strategy. Figures \ref{Fig1} and \ref{Fig2} show the recommended timeline and scenarios for different budget scenarios. %

{%
\vskip 0.5 \baselineskip
\noindent\hypertarget{rec1}{{\bf Recommendation 1}}: As the highest priority independent of the budget scenarios, complete construction projects and support operations of ongoing experiments and research to enable maximum science. We reaffirm the previous P5 recommendations on major initiatives:

HL-LHC (including the ATLAS and CMS detectors, as well as the Accelerator Upgrade Project) to start addressing why the Higgs boson condensed in the universe (reveal the secrets of the Higgs boson, \href{https://www.usparticlephysics.org/2023-p5-report/decipher-the-quantum-realm.html#32reveal-the-secrets-of-the-higgs-boson}{section 3.2}), to search for direct evidence for new particles \href{https://www.usparticlephysics.org/2023-p5-report/explore-new-paradigms-in-physics.html#51search-for-direct-evidence-of-new-particles}{(section 5.1)}, to pursue quantum imprints of new phenomena \href{https://www.usparticlephysics.org/2023-p5-report/explore-new-paradigms-in-physics.html#52pursue-quantum-imprints-of-new-phenomena}{(section 5.2)}, and to determine the nature of dark matter \href{https://www.usparticlephysics.org/2023-p5-report/illuminate-the-invisible-universe.html#41determine-the-nature-of-dark-matter}{(section 4.1)}.

The first phase of DUNE and PIP-II to open an era of precision neutrino measurements that include the determination of the mass ordering among neutrinos. Knowledge of this fundamental property is a crucial input to cosmology and nuclear science (elucidate the mysteries of neutrinos, \href{https://www.usparticlephysics.org/2023-p5-report/decipher-the-quantum-realm.html#31elucidate-the-mysteries-of-neutrinos}{section 3.1}).

The Vera C. Rubin Observatory to carry out the Legacy Survey of Space and Time (LSST), and the LSST Dark Energy Science Collaboration, to understand what drives cosmic evolution (\href{https://www.usparticlephysics.org/2023-p5-report/illuminate-the-invisible-universe.html#42understand-what-drives-cosmic-evolution}{section 4.2}). In addition, we recommend continued support for the following ongoing experiments at the medium scale (project costs $>$\$50M for DOE and $>$\$4M for NSF), including completion of construction, operations and research:
NOvA, SBN, T2K, and IceCube (elucidate the mysteries of neutrinos, \href{https://www.usparticlephysics.org/2023-p5-report/decipher-the-quantum-realm.html#31elucidate-the-mysteries-of-neutrinos}{section 3.1}).
DarkSide-20k, LZ, SuperCDMS, and XENONnT (determine the nature of dark matter, \href{https://www.usparticlephysics.org/2023-p5-report/illuminate-the-invisible-universe.html#41determine-the-nature-of-dark-matter}{section 4.1}).
DESI (understand what drives cosmic evolution, \href{https://www.usparticlephysics.org/2023-p5-report/illuminate-the-invisible-universe.html#42understand-what-drives-cosmic-evolution}{section 4.2}).
Belle II, LHCb, and Mu2e (pursue quantum imprints of new phenomena, \href{https://www.usparticlephysics.org/2023-p5-report/explore-new-paradigms-in-physics.html#52pursue-quantum-imprints-of-new-phenomena}{section 5.2}).

\vskip 0.5 \baselineskip
\hypertarget{rec2}{{\noindent\bf Recommendation 2}}: Construct a portfolio of major projects that collectively study nearly all fundamental constituents of our universe and their interactions, as well as how those interactions determine both the cosmic past and future.
These projects have the potential to transcend and transform our current paradigms. They inspire collaboration and international cooperation in advancing the frontiers of human knowledge. Plan and start the following major initiatives in order of priority from highest to lowest:

CMB-S4, which looks back at the earliest moments of the universe to probe physics at the highest energy scales. It is critical to install telescopes at and observe from both the South Pole and Chile sites to achieve the science goals (\href{https://www.usparticlephysics.org/2023-p5-report/illuminate-the-invisible-universe.html#42understand-what-drives-cosmic-evolution}{section 4.2}).

A re-envisioned second phase of DUNE with an early implementation of an enhanced 2.1 MW beam—ACE-MIRT, a third far detector, and an upgraded near-detector complex as the definitive long-baseline neutrino oscillation experiment of its kind (\href{https://www.usparticlephysics.org/2023-p5-report/decipher-the-quantum-realm.html#31elucidate-the-mysteries-of-neutrinos}{section 3.1}).

An offshore Higgs factory, realized in collaboration with international partners, in order to reveal the secrets of the Higgs boson. The current designs [as of 2023] %
of FCC-ee and ILC meet our scientific requirements. The US should actively engage in feasibility and design studies. Once a specific project is deemed feasible and well-defined (see also \hyperlink{rec6}{Recommendation 6}), the US should aim for a contribution at funding levels commensurate with that of the US involvement in the LHC and HL-LHC, while maintaining a healthy US onshore program in particle physics (\href{https://www.usparticlephysics.org/2023-p5-report/decipher-the-quantum-realm.html#32reveal-the-secrets-of-the-higgs-boson}{section 3.2}).

An ultimate Generation 3 (G3) dark matter direct detection experiment reaching the neutrino fog, in coordination with international partners and preferably sited in the US (\href{https://www.usparticlephysics.org/2023-p5-report/illuminate-the-invisible-universe.html#41determine-the-nature-of-dark-matter}{section 4.1}).
IceCube Gen-2, for study of neutrino properties complementary to DUNE and for indirect detection of dark matter covering higher mass ranges, using non-beam neutrinos as a tool. (\href{https://www.usparticlephysics.org/2023-p5-report/illuminate-the-invisible-universe.html#41determine-the-nature-of-dark-matter}{section 4.1}).

See Figure~\ref{Fig2} for estimated/intended cost bracket for various projects in {\it US accounting}\/.

\vskip 0.5 \baselineskip
\hypertarget{rec3}{{\noindent\bf Recommendation 3}}: Create an improved balance between small-, medium-, and large-scale projects to open new scientific opportunities and maximize their results, enhance workforce development, promote creativity, and compete on the world stage.

To achieve this balance across all project sizes we recommend the following:
Implement a new small-project portfolio at DOE, Advancing Science and Technology through Agile Experiments (ASTAE), across science themes in particle physics with a competitive program and recurring funding opportunity announcements. This program should start with the construction of experiments from the Dark Matter New Initiatives (DMNI) by DOE-HEP (\href{https://www.usparticlephysics.org/2023-p5-report/investing-in-the-future-of-science-and-technology.html}{section 6.2}).
Continue Mid-Scale Research Infrastructure (MSRI) and Major Research Instrumentation (MRI) programs as a critical component of the NSF research and project portfolio.
Support DESI-II for cosmic evolution, LHCb upgrade II and Belle II upgrade for quantum imprints, and US contributions to the global CTA Observatory for dark matter (sections \href{https://www.usparticlephysics.org/2023-p5-report/illuminate-the-invisible-universe.html#42understand-what-drives-cosmic-evolution}{4.2}, \href{https://www.usparticlephysics.org/2023-p5-report/explore-new-paradigms-in-physics.html#52pursue-quantum-imprints-of-new-phenomena}{5.2}, and \href{https://www.usparticlephysics.org/2023-p5-report/illuminate-the-invisible-universe.html#41determine-the-nature-of-dark-matter}{4.1}).
The Belle II recommendation includes contributions towards the Super-KEKB accelerator.

\vskip 0.5 \baselineskip
\hypertarget{rec4}{{\noindent\bf Recommendation 4}}: Support a comprehensive effort to develop the resources—theoretical, computational, and technological—essential to our 20-year vision for the field. This includes an aggressive R\&D program that, while technologically challenging, could yield revolutionary accelerator designs that chart a realistic path to a 10 TeV parton center-of-mass (pCM) collider.

Investing in the future of the field to fulfill this vision requires the following:
Support vigorous R\&D toward a cost-effective 10 TeV pCM collider based on proton, muon, or possible wakefield technologies, including an evaluation of options for US siting of such a machine, with a goal of being ready to build major test facilities and demonstrator facilities within the next 10 years (sections \href{https://www.usparticlephysics.org/2023-p5-report/decipher-the-quantum-realm.html#32reveal-the-secrets-of-the-higgs-boson}{3.2}, \href{https://www.usparticlephysics.org/2023-p5-report/explore-new-paradigms-in-physics.html#51search-for-direct-evidence-of-new-particles}{5.1}, \href{https://www.usparticlephysics.org/2023-p5-report/investing-in-the-future-of-science-and-technology.html#65collider-rd}{6.5}, and \hyperlink{rec6}{Recommendation 6}).
Enhance research in theory to propel innovation, maximize scientific impact of investments in experiments, and expand our understanding of the universe (\href{https://www.usparticlephysics.org/2023-p5-report/investing-in-the-future-of-science-and-technology.html#61theory}{section 6.1}).
Expand the General Accelerator R\&D (GARD) program within HEP, including stewardship (\href{https://www.usparticlephysics.org/2023-p5-report/investing-in-the-future-of-science-and-technology.html#64particle-accelerators-and-rd}{section 6.4}).
Invest in R\&D in instrumentation to develop innovative scientific tools (\href{https://www.usparticlephysics.org/2023-p5-report/investing-in-the-future-of-science-and-technology.html#63detector-instrumentation}{section 6.3}).
Conduct R\&D efforts to define and enable new projects in the next decade, including detectors for an $e^+e^-$ Higgs factory and 10 TeV pCM collider, Spec-S5, DUNE FD4, Mu2e-II, Advanced Muon Facility, and line intensity mapping (sections \href{https://www.usparticlephysics.org/2023-p5-report/decipher-the-quantum-realm.html#31elucidate-the-mysteries-of-neutrinos}{3.1}, \href{https://www.usparticlephysics.org/2023-p5-report/decipher-the-quantum-realm.html#32reveal-the-secrets-of-the-higgs-boson}{3.2}, \href{https://www.usparticlephysics.org/2023-p5-report/illuminate-the-invisible-universe.html#42understand-what-drives-cosmic-evolution}{4.2}, \href{https://www.usparticlephysics.org/2023-p5-report/explore-new-paradigms-in-physics.html#51search-for-direct-evidence-of-new-particles}{5.1}, \href{https://www.usparticlephysics.org/2023-p5-report/explore-new-paradigms-in-physics.html#52pursue-quantum-imprints-of-new-phenomena}{5.2}, and \href{https://www.usparticlephysics.org/2023-p5-report/investing-in-the-future-of-science-and-technology.html#63detector-instrumentation}{6.3}).
Support key cyberinfrastructure components such as shared software tools and a sustained R\&D effort in computing, to fully exploit emerging technologies for projects. Prioritize computing and novel data analysis techniques for maximizing science across the entire field (\href{https://www.usparticlephysics.org/2023-p5-report/investing-in-the-future-of-science-and-technology.html#67software-computing-and-data-science}{section 6.7}).
Develop plans for improving the Fermilab accelerator complex that are consistent with the long-term vision of this report including neutrinos, flavor, and a 10 TeV pCM collider (\href{https://www.usparticlephysics.org/2023-p5-report/investing-in-the-future-of-science-and-technology.html#66facilities-and-infrastructure}{section 6.6}).

\vskip 0.5\baselineskip
\hypertarget{rec5}{{\noindent\bf Recommendation 5}}: We omit this recommendation as it mainly discusses workforce development in the US.%
\vskip 0.5\baselineskip
\hypertarget{rec6}{{\noindent\bf Recommendation 6}}: Convene a targeted panel with broad membership across particle physics later this decade that makes decisions on the US accelerator-based program at the time when major decisions concerning an offshore Higgs factory are expected, and/or significant adjustments within the accelerator-based R\&D portfolio are likely to be needed. A plan for the Fermilab accelerator complex consistent with the long-term vision in this report should also be reviewed.

The panel would consider the following:

\begin{itemize}
\setlength\itemsep{0em}
\item The level and nature of US contribution in a specific Higgs factory including an evaluation of the associated schedule, budget, and risks once crucial information becomes available.
\item Mid- and large-scale test and demonstrator facilities in the accelerator and collider R\&D portfolios.

\item A plan for the evolution of the Fermilab accelerator complex consistent with the long- term vision in this report, which may commence construction in the event of a more favorable budget situation.
\end{itemize}
}
\vskip -0.5 \parskip
The P5 report also has more specific ``Area Recommendations'' that we do not include here.

\vspace{-1 em}
\subsection*{Developments since the 2023 P5 report}

Since the approval  of the P5 report, the funding agencies and US community have been actively implementing the recommendations. Notable implementation developments include:
\begin{itemize}
\setlength\itemsep{0em}

    \item DOE and NSF created the \href{https://higgsfactory.slac.stanford.edu/about}{Higgs Factory Coordination Consortium} (HFCC). Inspired by P5 \hyperlink{rec2}{Recommendation 2}, %
    it is charged to provide strategic direction and leadership for the U.S. community to engage, shape, and thereby advance the development of the physics, experiment, and detector (PED) and accelerator (A) programs for a potential future Higgs factory; and to ensure cooperation with our partners in the international program. They have convened a series of workshops over the past year.
    The HFCC is submitting a \href{https://indico.global/event/14242/contributions/122916/attachments/56846/110671/HFCC%20White%20Paper%20for%20ESG.pdf}{separate white paper}.
    \item Later in this decade, the funding agencies intend to form  dedicated panels to implement \hyperlink{rec6}{Recommendation 6}. These panels will determine the US commitment to a Higgs factory, iterate the roadmap for technical R\&D and develop long-term strategies for the Fermilab accelerator complex. 
    \item The formation of a \href{https://www.muoncollider.us}{US Muon Collider Collaboration} has been  launched. It is meant to work closely with the \href{https://muoncollider.web.cern.ch}{International Muon Collider Collaboration} hosted by CERN. The USMCC are submitting a \href{https://indico.fnal.gov/event/68182/attachments/185111/257117/ESPPU_USMCC_draft_v7.pdf}{separate white paper}. 
    \item A \href{https://arxiv.org/abs/2503.20214}{10 TeV Wakefield Collider Design Study} has been formed in response to P5 Recommendations 4 and 6. It was launched by scientists at US national laboratories, but has quickly developed into an international effort with strong engagement from European partners.
    \item The NSF has announced that it does not intend to move the CMB-S4 proposal to the next stage citing concerns about the deteriorating infrastructure at the South Pole station. DOE has instructed the CMB-S4 collaboration to present a revised proposal that does not include a South Pole site. At the same time, the community is advocating for improvements to South Pole infrastructure both for CMB-S4 and Ice-Cube. 
    \item There is   active discussion of the potential host country for the G3 dark matter direct detection experiment. DOE is planning  a selection process between xenon and argon. Coordination with the European community is crucial in this area.
    
    \item DOE intends to implement an ASTAE funding line to enable small-scale agile experiments starting with the five projects in the Dark Matter New Initiative (DMNI). TESSERACT is already moving forward thanks to a French commitment. %
    \item DOE-funded technology research programs for detectors and accelerators are developing roadmaps that embrace the R\&D priorities identified by the P5 report.  
    The US detector R\&D Collaborations, through the \href{https://cpad-dpf.org}{Coordinating Panel on Advanced Detectors} (CPAD), are building strong connections with the ECFA Detector R\&D collaborations. 
    \item US national laboratories are directly supporting the P5-identified R\&D priorities. %
    \item There is an ongoing study led by the National Academy of Sciences ``\href{https://www.nationalacademies.org/our-work/elementary-particle-physics-progress-and-promise}{Elementary Particle Physics: Progress and Promise}'' looking into the longer-term future of the field. The report is expected later in 2025.

\end{itemize}

The current budget for particle physics is unfortunately near the less favorable funding scenario. That scenario  would limit the implementation of the P5 recommendations beyond the current commitments.

In all of the above, the US community is looking forward to engagement of the European community in US-hosted initiatives and to continued collaboration on European-sited projects.

\vspace{-1 em}
\subsection*{The broader program - connections to astrophysics and nuclear physics. }

\subsubsection*{Astronomy and Astrophysics}

We note that, in the US, the particle physics program includes aspects of cosmology and astroparticle physics to understand the nature of Inflation,  Dark Energy and Dark Matter. 
Experiments in these areas are among the top priorities recommended by P5. %

The US community in astronomy and astrophysics has also gone through its study called \href{https://www.nationalacademies.org/our-work/decadal-survey-on-astronomy-and-astrophysics-2020-astro2020}{Astro2020} led by the National Academy. It resulted in the report ``\href{https://nap.nationalacademies.org/catalog/26141/pathways-to-discovery-in-astronomy-and-astrophysics-for-the-2020s}{Pathways to Discovery in Astronomy and Astrophysics for the 2020s}.'' It ranked CMB-S4 and IceCube Gen2 highly, consistent with the P5 recommendations.

\subsubsection*{Nuclear Physics}

In the US, nuclear physics is funded separately from particle physics but our scientific interests overlap considerably and we use similar facilities and experimental techniques. 

In 2022, the DOE Office of Science and the NSF requested a 
long-range plan that would provide a framework for coordinated advancement of the
US nuclear science research programs over the next decade. A working group was charged with drafting a report, and the US nuclear physics community organized three town hall meetings -- one for each broad subdiscipline within the nuclear physics community -- in late 2022 to collect feedback from a larger subset of the community and to identify opportunities and challenges.   
The 2023 Long Range Plan for Nuclear Science (\href{https://nuclearsciencefuture.org/wp-content/uploads/2024/03/23-G06476-2024-LRP-8.5x11-pcg-v1.5-3.14.24.pdf}{LRP}) was released in the Fall of 2023. Some of its recommendations and discussions are directly relevant to the ESPP and are summarized here. Similar to our discussion of the P5 recommendations, we restrict the presentation to, for the most part, the text that is presented in the  \href{https://nuclearsciencefuture.org/wp-content/uploads/2024/03/23-G06476-2024-LRP-8.5x11-pcg-v1.5-3.14.24.pdf}{LRP}, to which we refer for the broader context. 

According to the LRP, nuclear science addresses some of the outstanding
challenges to modern physics, including the properties and limits of matter, the forces of nature, and the evolution of the universe:
\begin{itemize}
\setlength\itemsep{0em}
\item How do quarks and gluons make up protons, neutrons, and, ultimately, atomic nuclei?
\item How do the rich patterns observed in the structure and reactions of nuclei emerge from
the interactions between neutrons and protons?
\item What are the nuclear processes that drive the birth, life, and death of stars?
\item How do we use atomic nuclei to uncover physics
beyond the Standard Model?
\end{itemize}

The pursuit of the first and fourth of these fundamental physics questions overlaps quite directly with the goals of particle physics. The following recommendations are directly relevant to the ESPP:
\begin{enumerate}
 \item
The highest priority of the nuclear science community is to capitalize on the extraordinary opportunities for scientific discovery made possible by the substantial and sustained investments of the United States. We must draw on the talents of all in the nation to achieve this goal. 
This includes completing the RHIC science program.
\item As the highest priority for new experiment construction, we recommend that the United States
lead an international consortium that will undertake a neutrinoless double beta decay campaign,
featuring the expeditious construction of ton-scale experiments, using different isotopes and complementary techniques.
\item We recommend the expeditious completion of the Electron--Ion Collider
(EIC) as the highest priority for facility construction.
\end{enumerate}
    
The two highest-priority new initiatives -- a campaign to improve our sensitivity to neutrinoless double-beta decay and the EIC -- share scientific motivation and techniques with particle physics. Both rely heavily on cooperation and collaboration with European Institutions.

\vspace{-1 em}
\subsection*{US-European Partnerships}

Particle, nuclear and accelerator science in the US and Europe are intertwined by shared facilities, technologies and scientific goals. The enabling power of international partnership was emphasized in a 2023 HEPAP subpanel report, “The Path to Global Discovery – U.S. Leadership and Partnership in Particle Physics.” This report noted that “hosting unique world-class facilities at home and partnering in high-priority facilities hosted elsewhere” are both “essential components of an achievable global vision for the field”. Here we quote only three recommendations in sections 3.3 and 3.4 from that report:%
\begin{itemize}
\setlength\itemsep{0em}
\item        Continue support for and actively seek engagement with international collaborations and partnerships of all sizes.
\item        The U.S. particle physics program should 1) strive to engage as partners in the construction and operation of major future particle physics accelerator facilities constructed outside the U.S. and 2) actively seek international partners to engage in the construction and operation of major future particle accelerator facilities constructed in the U.S. 
\item        Implement structures for hosting strong international collaborations, act with timeliness, consistently meet obligations, and facilitate open communication with partners
\end{itemize}

In addition, we have identified numerous examples of cooperation towards shared goals.  An incomplete list includes:

\begin{itemize}
\setlength\itemsep{0em}
\item  Eight US national laboratories and over 80 universities have participated in LHC experiments since the 1990's,  contributing both to the machine and the experiments for the LHC and now the HL-LHC. US scientists participate mainly ATLAS and CMS (making up 18\% and 30\% of the collaborations respectively), with smaller participation in LHCb and ALICE. The US community remains  committed to continue these efforts through the HL-LHC era.
\item In addition to multiple contributions to experiment upgrades for the HL-LHC, the US is contributing superconducting quadrupole magnets and RF systems to the accelerator complex.
\item The DOE and NSF have also launched initiatives to develop advanced computing and software techniques to cope with the large volume of data from the HL-LHC and DUNE.  
\item Building on a long history of European participation in  experiments at Fermilab, the DUNE/LBNF neutrino experiment has substantial European involvement.
47\% of DUNE's scientific collaborators are from the US, 39\% from Europe, including CERN, with the remaining 14\% from Asia, Africa, Canada, and Latin America. Close to 50\% of the detector components and computing resources are being contributed by European institutions. CERN is making major contributions to DUNE through the Neutrino Platform and provision of far detector cryostats for LBNF.  
The DUNE collaboration is submitting \href{https://indico.global/event/14242/timetable/#4-duneneutrino-experiments}{four white papers}.

 \item The PIP-II accelerator upgrade project at Fermilab depends upon accelerator components from France, India, Italy, Poland, and the UK. The LBNF/PIP-II accelerator complex has completed construction of conventional facilities and is beginning delivery and installation of cryogenic and accelerator components.
 \item In response to P5, DOE \href{https://science.osti.gov/-/media/hep/hepap/pdf/Meetings/2024/DOE-HEP-P5-G-Rameika-V3_2.pdf}{fully supports} the implementation of the first phase of DUNE and PIP-II as one of their highest priority projects.  DOE also supports the second phase of DUNE with a focus on ACE-MIRT, a third far detector and a more capable near detector.
\item The Electron Ion Collider project builds on a long history of international collaboration in nuclear physics at multiple US and European laboratories. There is strong participation in the detector collaboration and engagement on accelerator concepts and technologies that are relevant both for the EIC and FCC development.
 \item There is broad cross-Atlantic participation in smaller experiments as well. European collaborators play a major role in US experiments such as \href{https://muon-g-2.fnal.gov/collaboration.html}{$g-2$}, \href{https://mu2e.fnal.gov}{Mu2e}, \href{https://lz.lbl.gov}{LZ}, \href{https://sbn-nd.fnal.gov}{SBND}, and \href{https://icarus.fnal.gov}{ICARUS} while US scientists collaborate at CERN on multiple non-LHC efforts.  %
 \item US accelerator physicists engaged strongly with the development of the \href{https://e-publishing.cern.ch/index.php/CYRM/issue/view/146}{European Strategy for Particle Physics -- Accelerator R\&D Roadmap}, which was prepared by the European Laboratory Director's Group after the previous European Strategy Update.  There are also strong connections in Detector R\&D between US and European efforts.

\end{itemize}

\vspace{-1 em}
\subsection*{Response to \href{https://ecfa.web.cern.ch/ecfa-guidelines-inputs-national-hep-communities-european-strategy-particle-physics-0}{ECFA questions}}

\href{https://indico.global/event/14242/timetable/}{US white papers} and the \href{https://indico.global/event/13796/}{DPF/DPB community forum} informed 
our responses, which are italicized below.
 
3) Questions to be considered by countries/regions when forming and submitting their “national input” to the ESPP:
\renewcommand{\labelenumi}{\alph{enumi})}
\begin{enumerate}
\item Which is the preferred next major/flagship collider project for CERN?\\ 
{\it The P5 report \hyperlink{rec2}{recommended} that the preferred next collider should be a Higgs factory. P5  concluded that such a facility could  not be hosted by the US within the budget guidelines from the agencies. P5 identified both FCC-ee and ILC as projects that could meet the scientific requirements without specifying their locations. In preparation for the ESPP update, the US funding agencies (DOE and NSF) \href{https://higgsfactory.slac.stanford.edu/about}{commissioned} a focused Higgs Factory Coordination Consortium whose report is being submitted as a \href{https://indico.global/event/14242/contributions/122916/attachments/56846/110671/HFCC%20White%20Paper%20for%20ESG.pdf}{separate document}. 
That report concludes: \\

\vskip -1 em

''The U.S. is enthusiastic for a Higgs Factory as the next major collider and strongly supports FCC-ee ... if it is chosen as the next major
research infrastructure project at CERN. ...
The U.S. would also support an LC if the CERN
Council approves such a project in a timely manner.''%
}
\item What are the most important elements in the response to (a)?\\
{\it Physics potential is the primary motivation, please see the \href{https://indico.global/event/14242/contributions/122916/attachments/56846/110671/HFCC%20White%20Paper%20for%20ESG.pdf}{HFCC} white paper for a more detailed response.} 

\item Should CERN/Europe proceed with the preferred option set out in (a) or should alternative options be
considered?\\
{\it \hyperlink{rec6}{Recommendation 6} from P5 is that the US should revisit this question in light of international developments later in this decade.}
\item Beyond the preferred option in (a), what other accelerator R\&D topics (e.g. high field magnets, RF
technology, alternative accelerators/colliders) should be pursued in parallel?\\
{\it The US is working  closely with our European colleagues to develop new technologies. Development of high-field magnets, RF technology, and collider concepts that can achieve a 10 TeV pCM scale are high priorities for the US. A \href{https://indico.fnal.gov/event/68182/attachments/185111/257117/ESPPU_USMCC_draft_v7.pdf}{US white paper} focused on R\&D for a future muon collider is being submitted as a separate document. These priorities will be revisited later this decade.}%
\item What is the prioritized list of alternative options if the preferred option set out in (a) is not feasible?\\
{\it This question should  be answered later in this decade when the global situation is better understood.}
\item What are the most important elements in the response to (e)? \\
{\it This question should  be answered later in this decade when the global situation is better understood.}
\end{enumerate}
\vskip -2 pt
4) The remit given to the ESG also specifies that “The Strategy update should also indicate areas of priority for exploration complementary to colliders and for other experiments to be considered at CERN and at other laboratories in Europe, as well as for participation in projects outside Europe.” %
\begin{enumerate}
\setlength\itemsep{0em}
\item What other areas of physics should be pursued, and with what relative priority?

\item What are the most important elements in the response to 4a)? (The set of considerations in 3b should be used).  

\item  To what extent should CERN participate in nuclear physics, astroparticle physics or other areas of science, while keeping in mind and adhering to the CERN Convention? Please use the current level and form of activity as the baseline for comparisons.
\end{enumerate}

{\it We provide a combined answer to questions 4a), b), and c):
As noted in P5 recommendations \hyperlink{rec2}{2} and \hyperlink{rec3}{3}, major efforts in neutrino physics, dark matter, and cosmic microwave background (CMB) should be aggressively pursued, in parallel to a suite of smaller experiments, both in the US and in Europe.  In addition to the particle physics recommendations, the EIC and double-beta decay are top priorities for US nuclear science.}
 
\vspace{-1 em}
\subsection*{Conclusion}
In this document, we have summarized the results of the US community prioritization process.  The P5 process was the result of broad community input followed by a  rigorous prioritization process,  only retaining the most compelling projects that were consistent with known budget and time constraints.  Those priorities were then explicitly endorsed by a majority of the US particle physics community. Priorities for CERN do remain a dominantly European decision but
our mutual goal should be for a timely realization of high-impact science with an optimization of worldwide resources.

\end{document}